\documentclass[prl,twocolumn,showpacs,amsmath,amssymb]{revtex4}

\usepackage{graphicx}
\usepackage{dcolumn}
\usepackage{bm}
\usepackage{epsfig}

\begin{document}

\newcommand{\ep}{\varepsilon}
\newcommand{\up}{\uparrow}
\newcommand{\dn}{\downarrow}
\newcommand{\vectg}[1]{\mbox{\boldmath ${#1}$}}
\newcommand{\vect}[1]{{\bf #1}}

\title{Extrinsic Entwined with Intrinsic Spin Hall Effect in Disordered Mesoscopic Bars}

\author{Branislav K. Nikoli\' c}
\author{Liviu P. Z\^ arbo}
\affiliation{Department of Physics and Astronomy, University
of Delaware, Newark, DE 19716-2570, USA}

\begin{abstract}
We show that pure spin Hall current, flowing out of a four-terminal phase-coherent two-dimensional electron gas (2DEG) within inversion asymmetric semiconductor heterostructure, contains contributions from  both the extrinsic mechanisms (spin-orbit dependent scattering off impurities) and the intrinsic ones  (due to the Rashba coupling). While the extrinsic contribution vanishes in the weakly and strongly disordered limits, and the intrinsic one dominates in the quasiballistic limit, in the crossover transport regime the spin Hall conductance, exhibiting sample-to-sample large fluctuations and sign change, is not simply reducible  
to either of the two mechanisms, which can be relevant for interpretation of experiments on dirty 2DEGs [V.~Sih {\em et al.}, Nature Phys. {\bf 1}, 31 (2005)].
\end{abstract}

\maketitle

{\it Introduction}.---Recent experimental discovery of the spin Hall effect in three-dimensional $n$-type semiconductors~\cite{kato2004a} and two-dimensional hole~\cite{wunderlich2005a} or electron gases~\cite{sih2005a} provides deep insight into the relativistic effects in solids which manifest 
through spin-orbit (SO) couplings~\cite{winkler_book}, as well as into the nature of spin accumulation 
and spin fluxes in systems whose physics is governed by the SO interactions. The flow  of conventional unpolarized charge current in the longitudinal  direction through such systems leads to spin flux in the transverse direction which deposits nonequilibrium  spin accumulation on the lateral 
sample edges, as observed in experiments~\cite{kato2004a,wunderlich2005a,sih2005a}. Moreover, the 
anticipated experiments should demonstrate flow of spin into the transverse electrodes attached at 
those edges~\cite{nikolic_images}, thereby offering a semiconductor analog of the Stern-Gerlach 
device as a source of spin currents or flying spin qubits where spatial separation of  spin-$\uparrow$ 
and spin-$\downarrow$ electrons does not require any cumbersome external magnetic fields.

However, the effect observed in electron systems~\cite{kato2004a,sih2005a} is rather small, and 
its magnitude cannot be tuned easily since it is determined by material parameters which govern the 
SO dependent scattering off impurities~\cite{dyakonov1971a,engel2005a,tse2006a} deflecting a beam 
of spin-$\uparrow$ (spin-$\downarrow$) electrons predominantly to the right (left). 
In contrast to such extrinsically (i.e., impurity) induced  spin Hall effect, recent theories have 
argued for several orders of magnitude larger spin Hall currents due to the intrinsic SO couplings 
capable of spin-splitting the energy bands and inducing transverse SO forces~\cite{nikolic_images} 
in bulk semiconductors in the clean limit~\cite{murakami2003a,sinova2004a}, as well as in ballistic 
mesoscopic multiterminal nanostructures made of such materials~\cite{nikolic_images,nikolic_mesoshe}. Such mechanisms were invoked to explain~\cite{nomura2005b} observation of two orders of magnitude larger spin 
Hall accumulation in 2D hole gases~\cite{wunderlich2005a}, whose magnitude can still be affected by 
the effects of the disorder~\cite{chen2005a}. For example, in infinite 2DEGs 
with linear in momentum SO coupling arbitrarily small amount of impurities completely suppresses  the 
spin Hall effect~\cite{inoue2004a}, while increasing disorder only gradually diminishes the signatures  
of the spin Hall effect in mesoscopic 2DEG structures~\cite{nikolic_mesoshe,sheng2005a}.

Despite these advances, the {\em intrinsic vs. extrinsic} debate on the origin of the detected signatures of 
the spin Hall effect persists~\cite{zhang2005a,niu2006a}, closely mimicking several decades old discourse on the Karplus-Luttinger intrinsic  vs. skew-scattering + side-jump (i.e., shift of the scattered wave packet)~\cite{nozieres1973a} possible explanations of the anomalous Hall effect observed  in ferromagnetic systems~\cite{ahe_experiment}. Surprisingly enough, {\em no theory} has been constructed to describe spin current in the transition from quasiballistic to dirty transport regime in multiterminal Hall bars employed in experiments where, in principle, both extrinsic and intrinsic mechanisms could be equally important. Also, at low temperatures, quantum  corrections to the semiclassical description~\cite{engel2005a,tse2006a} of the extrinsic spin Hall effect, such as the mesoscopic fluctuations and localization effects due to the interplay of randomness and quantum coherence, can play an essential role for interpreting experiments on meso- and nano-scale samples. 

The demand for  such theory is further provoked by very recent spin Hall experiment on dirty two-terminal 
2DEG within AlGaAs quantum well which accommodates both the SO scattering off impurities and a small Rashba SO coupling due to weak structural inversion asymmetry. The latter could give rise to an intrinsic 
contribution of a similar magnitude as the extrinsic one for not too small ratio  $\Delta_{\rm SO} \tau/\hbar$~\cite{nikolic_mesoshe,sheng2005a} ($\Delta_{\rm SO}$ is the spin-splitting of quasiparticle energies and $\hbar/\tau$ is the disorder induced broadening of energy levels due to finite 
transport scattering time $\tau$)---the parameters measured in Ref.~\cite{sih2005a} place its Hall bar  
into the regime $\Delta_{\rm SO} \tau/\hbar \sim 10^{-2}$. 

Here we construct a {\em quantum transport} theory of the spin Hall fluxes driven by a generic 
Hamiltonian 
\begin{eqnarray}\label{eq:ahe}
\hat{H} & = & \frac{\hat{p}_x^2+\hat{p}_y^2}{2m^*} + V_{\rm confine}(y) + V_{\rm disorder}(x,y) \nonumber \\ && -\frac{g\mu_B}{2} \hat{\vectg{\sigma}} \cdot {\bf B}({\bf p}) + \lambda \left(\hat{\vectg{\sigma}}\times\hat{\vect{p}} \right) \cdot \nabla V_{\rm disorder}(x,y),
\end{eqnarray}
which encompasses both the intrinsic (third term) and the extrinsic (fourth term) SO coupling-induced effects. 
\begin{figure}
\centerline{\psfig{file=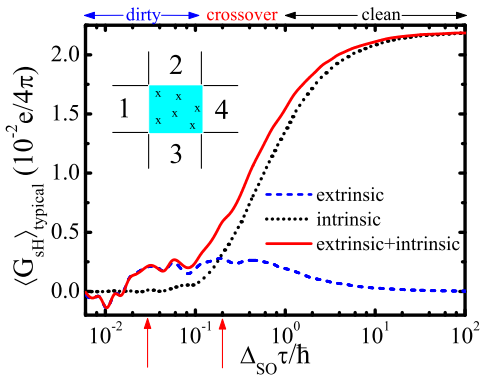,scale=0.6,angle=0}}
\caption{(Color online) The disorder-averaged {\em typical} spin Hall conductance of 
a four-terminal phase-coherent 2DEG nanostructure of the size $100a \times 100a$ 
($a \simeq 2$ nm), which is governed by both the SO scattering off impurities 
(setting the transport scattering time $\tau$) of strength $\lambda_{\rm SO}$ and the intrinsic Rashba SO coupling responsible for the spin-splitting  $\Delta_{\rm SO}=4at_{\rm SO}k_F$. Following measured parameters of 2DEG in Ref.~\cite{sih2005a}, we use $\lambda_{\rm SO}=0.005$, $t_{\rm SO}=0$ for the ``extrinsic''; $\lambda_{\rm SO}=0$, $t_{\rm SO}=0.003t_{\rm o}$, for the ``intrinsic''; and $\lambda_{\rm SO} =0.005$, $t_{\rm SO}=0.003t_{\rm o}$ for the ``extrinsic + intrinsic'' spin Hall conductance $G_{sH} = I_2^s/(V_1-V_4)$ of the pure spin current $I_2^s$ in the transverse ideal lead 2.}\label{fig:she_intro}
\end{figure}
Within this framework, we obtain in Fig.~\ref{fig:she_intro} the spin Hall conductance in the transverse ideal (i.e., spin and charge interaction-free) leads of a four-terminal quantum-coherent Hall bar containing such 2DEG.  While the intrinsic SO coupling term has general Zeeman form  $-g\mu_B \hat{\vectg{\sigma}} \cdot {\bf B}({\bf p})/2$, its effective magnetic field ${\bf B}({\bf p})$ is momentum dependent and does not break the time-reversal invariance. For the Rashba coupling, which arises due to structural inversion asymmetry (of the 2DEG confining electric potential along the $z$-axis and differing  band discontinuities at the heterostructure quantum  well  interface~\cite{winkler_book}), ${\bf B}({\bf p})=-(2\alpha/g\mu_B)(\hat{\bf p} \times \hat{z})$ where $\hat{z}$ is the unit vector orthogonal to 2DEG. The corresponding spin-splitting of the energy bands at the Fermi level ($\hbar k_F$ is the Fermi momentum) 
is $\Delta_{\rm SO} = g \mu_B |{\bf B}({\bf p})|= 2 \alpha k_F$.

The Thomas term $\lambda \left(\hat{\vectg{\sigma}}\times\hat{\vect{p}} \right) \cdot \nabla V_{\rm dis}({\bf r})$ is a relativistic correction to the Pauli equation for spin-$\frac{1}{2}$ particle where $\lambda/\hbar=-\hbar^2/4m_0^2c^2 \approx -3.7 \times 10^{-6}$ \AA{}$^2$ ($c$ is velocity of light) for electron in vacuum with mass $m_0$. While this vacuum value would lead to hardly observable spin Hall effect, in solids $\lambda$ can be enormously renormalized by the band structure effects due to the strong crystal potential---for example, $\lambda/\hbar =5.3$ \AA{}$^2$ in GaAs~\cite{winkler_book} generates the extrinsic spin Hall conductivity $\sigma_{sH}^{\rm extr} = j_y^z/E_x$ whose recently computed value~\cite{engel2005a,tse2006a} could account for the magnitude of the observed spin Hall accumulation in dirty 3D electron systems of Ref.~\cite{kato2004a}. In the inversion symmetric systems ${\bf B}({\bf p}) \equiv 0$, so that pure spin current $j_y^z$ (carrying $z$-axis polarized spins along the $y$-axis) is driven solely by the Thomas term with $\lambda \neq 0$ in the presence of longitudinal external electric field $E_x$.

\begin{figure}
\centerline{\psfig{file=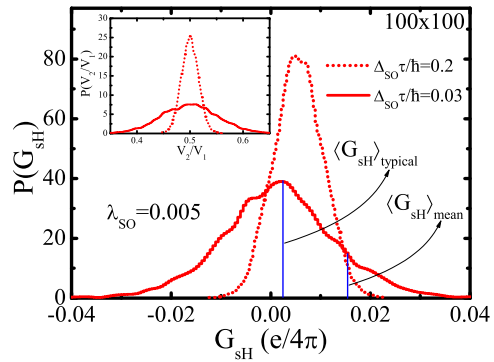,scale=0.6,angle=0}} 
\caption{(Color online) The full distribution function $P(G_{sH}$) of the spin 
Hall conductance $G_{sH}$ fluctuating from sample to sample in an ensemble of 
10000 impurity configurations within 2DEG with the extrinsic SO scattering  $\lambda_{\rm SO}=0.005$ 
and the intrinsic Rashba SO coupling $t_{\rm SO}=0.003t_{\rm o}$. The inset shows the corresponding distribution function of voltages $V_2$  which have to be applied to the top transverse lead (together with voltage $V_3$ on the bottom lead) to ensure that no charge current $I_2=I_3=0$ accompanies pure spin current $I_2^s$.}\label{fig:spin_ucf}
\end{figure}

The extrinsic spin Hall {\em conductivity} $\sigma_{sH}^{\rm extr} = \sigma_{sH}^{\rm SS}+\sigma_{sH}^{\rm SJ}$, which is the sum~\cite{engel2005a,tse2006a} of the skew-scattering $\sigma_{sH}^{\rm SS}<0$ and the side-jump $\sigma_{sH}^{\rm SJ}>0$ contributions, is a singular function as the disorder strength is decreased to zero since $\sigma_{sH}^{\rm SS} \sim \tau$ while  $\sigma_{sH}^{\rm SJ}$ is $\tau$-independent~\cite{engel2005a,tse2006a,nozieres1973a}. Therefore, in the clean 
limit this divergence is expected to be cut off at $\tau \sim \hbar/\Delta_{\rm SO}$, where $\Delta_{\rm SO}$ 
is due to the $k^3$ Dresselhaus term in the absence of Rashba coupling~\cite{engel2005a}.  In contrast to this unphysical behavior of bulk transport quantities, Fig.~\ref{fig:she_intro} shows that the  extrinsic spin Hall {\em conductance} of the four-probe mesoscopic bar, $G_{sH}^{\rm extr}=G_{sH}(\lambda\neq0,\alpha=0)$ goes to zero $G_{sH} \rightarrow 0$ in the clean limit $\Delta_{\rm SO} \gg \hbar/\tau$, as well as in the very ($k_F\ell \sim 1$) dirty limit $\Delta_{\rm SO} \ll \hbar/\tau$ where semiclassical theories break down. The full spin conductance is defined as $G_{sH}=I_2^s/(V_1-V_4)$ for transverse spin current $I_2^s=\frac{\hbar}{2e}(I_2^\uparrow -I_2^\downarrow)$ of $z$-polarized spins driven by the voltage bias $V_1-V_4$ [see inset of Fig.~\ref{fig:she_intro}].

Although we find that the disorder-averaged extrinsic spin Hall conductance $\langle G_{sH}^{\rm extr} 
\rangle = \langle G_{sH}(\lambda,\alpha \equiv 0) \rangle$ depends linearly on $\lambda$, as is the case of the extrinsic bulk spin Hall conductivity~\cite{tse2006a} $\sigma_{sH}^{\rm extr} \propto \lambda$, Fig.~\ref{fig:spin_ucf} demonstrates that in phase-coherent nanostructures one also has to take into account violent mesoscopic fluctuations on the top of semiclassical background---they affect the change of the sign of $G_{sH}$ from sample to sample and generate wide distribution of conductances $P(G_{sH})$. Thus, we use the typical value~\cite{dobro2003a} $\langle G_{sH} \rangle \equiv \langle G_{sH} \rangle_{\rm typical}$, at which  the corresponding full distribution function $P(G_{sH})$ reaches its maximum (see Fig.~\ref{fig:spin_ucf} for illustration), to characterize an ensemble of mesoscopic spin Hall bars differing from each other in the impurity configuration. Note that both $\langle G_{sH} \rangle_{\rm typical}$ and $ \langle G_{sH} \rangle_{\rm mean}$ are positive within the dirty regime of recent experiments~\cite{kato2004a,sih2005a}, meaning that the spin Hall current $I_2^s$ flows from the top to the bottom transverse lead because of spin-$\uparrow$ being deflected to the right, in accord with reported sign of lateral spin accumulation in two-terminal microstructures~\cite{kato2004a,sih2005a}.

{\it Extrinsic + intrinsic total spin Hall currents in terminals of mesoscopic bars}.---Theoretical analysis of the spin Hall effect in bulk systems yields  $\sigma_{sH}$ by computing (via the Kubo formula~\cite{murakami2003a,sinova2004a,inoue2004a,tse2006a} or kinetic equations~\cite{engel2005a})  
the pure spin current density $j_y^z$, as the expectation  value of some plausibly introduced spin current operator $\hat{j}_y^z$~\cite{niu2006a}, in linear response to external electric field $E_x$ penetrating an infinite semiconductor. This procedure is well-defined in inversion symmetric systems (such as $\lambda \neq 0$,  $\alpha=0$) where SO contribution to the impurity potential decays  fast and spin current is a conserved quantity~\cite{engel2005a}. However, non-conservation of spin in spin-split systems with ${\bf B}({\bf p}) \neq 0$ leads to  ambiguity in the definition of spin current and the corresponding $\sigma_{sH}$~\cite{niu2006a}. That is, the spin density relaxes not only due to the spin flux but also due 
to spin precession, thereby requiring careful treatment of boundary effects on the observed spin accumulation~\cite{galitski2006}.

To generate analogous phenomenology in multiterminal Hall bars we have to employ at least four external 
leads  where passing unpolarized charge current through the longitudinal electrodes generates spin current 
in the transverse electrodes~\cite{nikolic_images,nikolic_mesoshe}. The charge currents in the terminals are described by the same multiprobe Landauer-B\" uttiker scattering formulas employed for mesoscopic quantum Hall bars~\cite{ando2003}, $I_p = \sum_q G_{pq}(V_p - V_q)$, which relate current in probe $p$ to voltages $V_q$ in all probes attached to the sample via conductance coefficients $G_{pq}$ that intrinsically take into account all interfaces and boundaries.  To ensure that transverse spin Hall current is {\em pure}, one has to apply transverse voltages $V_2$ and $V_3$ obtained from these equations for $I_2=I_3=0$. The extension of the scattering theory of quantum transport to {\em total}~\cite{nikolic_images} spin currents in the terminals $I_p^s=\frac{\hbar}{2e}(I_p^\uparrow - I_p^\downarrow)$, which are always genuine {\em nonequilibrium} response  (in contrast to spin current density which can be  non-zero even in equilibrium~\cite{nikolic_images}) and {\em conserved} in the ideal [${\bf B}({\bf p}) \equiv 0$] leads, yields~\cite{nikolic_mesoshe}
\begin{equation} \label{eq:spinbuttiker}
I_p^s  = \frac{\hbar}{2e}\sum_{q \neq p} (G_{qp}^{\rm out} V_p - G_{pq}^{\rm in} V_q),
\end{equation} 
where $G_{pq}^{\rm in}=G_{pq}^{\uparrow\uparrow}+G_{pq}^{\uparrow\downarrow}-G_{pq}^{\downarrow\uparrow}-G_{pq}^{\downarrow\downarrow}$, $G_{qp}^{\rm out}=G_{qp}^{\uparrow\uparrow}+G_{qp}^{\downarrow\uparrow}-G_{qp}^{\uparrow\downarrow}-G_{qp}^{\downarrow\downarrow}$, and the standard charge conductance coefficients are $G_{pq}=G_{pq}^{\uparrow\uparrow}+G_{pq}^{\uparrow\downarrow}+G_{pq}^{\downarrow\uparrow}+G_{pq}^{\downarrow\downarrow}$. The spin-resolved conductances are obtained from the transmission  matrices ${\bf t}^{pq}$  between the leads $p$ and $q$ through the Landauer-type formula $G_{pq}^{\sigma \sigma^\prime}=\frac{e^2}{h} \sum_{i,j=1} |{\bf t}^{pq}_{ij,\sigma \sigma^\prime}|^2$, where $\sum_{i,j=1} |{\bf t}^{pq}_{ij,\sigma \sigma^\prime}|^2$ is the probability for spin-$\sigma^\prime$  electron incident in lead $q$ to be transmitted to lead $p$ as spin-$\sigma$ electron and $i,j$ label the transverse propagating modes in the leads. Solving Eq.~(\ref{eq:spinbuttiker}) for $I_2^s$ in the top transverse lead of the Hall 
 bar in Fig.~\ref{fig:she_intro} yields a general formula~\cite{nikolic_mesoshe}
\begin{equation} \label{eq:gsh_explicit}
G_{sH}=\frac{\hbar}{2e} \left[(G_{12}^{\rm out}+G_{32}^{\rm out}+ G_{42}^{\rm out})\frac{V_2}{V_1} - G_{23}^{\rm in}\frac{V_3}{V_1} - G_{21}^{\rm in} \right],
\end{equation}
where we choose the reference potential $V_4=0$ and the $z$-axis is selected as the spin quantization axis 
for $\uparrow$, $\downarrow$. 

The consequences of the scattering approach-based four-terminal spin Hall conductance formula Eq.~(\ref{eq:gsh_explicit}) can be worked out either by analytical means, such as the random matrix theory applicable to ``black-box'' disordered or chaotic ballistic structures, or by switching to nonperturbative  real$\otimes$spin space Green functions~\cite{nikolic_images} to take into account microscopic Hamiltonian which models the details of scattering and SO coupling effects in finite-size samples attached to many external probes. For this purpose, we represent the Rashba~\cite{nikolic_mesoshe} and the Thomas term~\cite{pareek2001} in Eq.~(\ref{eq:ahe}) in the local orbital basis
\begin{widetext}
\begin{eqnarray}\label{eq:tbh}
\hat{H} =  \sum_{{\bf m},\sigma} \varepsilon_{\bf m} \hat{c}_{{\bf
m}\sigma}^\dag\hat{c}_{{\bf m}\sigma}+\sum_{\langle {\bf
mm'}\rangle}  \sum_{\sigma\sigma'} \hat{c}_{{\bf m}\sigma}^\dag t_{\bf
mm'}^{\sigma\sigma'}\hat{c}_{{\bf m'}\sigma'}  
 -i  \lambda_{\rm SO} \sum_{{\bf m},\alpha\beta} \sum_{ij} \sum_{\nu\gamma} \epsilon_{ijz} \nu \gamma (\varepsilon_{{\bf m} + \gamma {\bf e}_j} - \varepsilon_{{\bf m}+\nu {\bf e}_i}) \hat{c}_{{\bf
m},\alpha}^\dag \hat{\sigma}^z_{\alpha\beta} \hat{c}_{{\bf m}+\nu{\bf e}_i+\gamma{\bf e}_j,\beta}.
\end{eqnarray}
\end{widetext}
The first term models isotropic short-range spin-independent static impurity potential where $\varepsilon_{\bf m} \in [-W/2,W/2]$ is a uniform random variable. The second, Rashba term, is a tight-binding type of Hamiltonian whose nearest-neighbor $\langle {\bf mm'}\rangle$ hopping is a  non-trivial $2 \times 2$ Hermitian  matrix  ${\bf t}_{\bf m'm}=({\bf t}_{\bf mm'})^\dagger$ in the spin space 
\begin{eqnarray}\label{eq:hopping}
{\bf t}_{\bf mm'}=\left\{
\begin{array}{cc}
-t_{\rm o}{\bf I}_s-it_{\rm SO}\hat{\sigma}_y &
({\bf m}={\bf m}'+{\bf e}_x)\\
-t_{\rm o}{\bf I}_s+it_{\rm SO}\hat{\sigma}_x &  ({\bf m}={\bf m}'+{\bf e}_y).
\end{array}\right.
\end{eqnarray}
The strength of the SO coupling is measured by the parameter $t_{\rm SO}=\alpha/2a$ (${\bf I}_s$ is the unit
$2 \times 2$ matrix in the spin space) which determines the spin-splitting of the band structure $\Delta_{\rm SO}=4at_{\rm SO}k_F$. A direct correspondence between  the continuous effective Hamiltonian Eq.~(\ref{eq:ahe})  and its lattice version Eq.~(\ref{eq:tbh}) is established  by selecting the Fermi energy ($E_F=-3.5t_{\rm o}$ in the rest of the paper) of the injected electrons to be close to the bottom of the band where tight-binding dispersion reduces to the parabolic one, and by using $t_{\rm o}=\hbar^2/(2 m^* a^2)$ for the orbital hopping  which yields the effective mass $m^*$ in the continuum limit.  The labels in the third (Thomas) term, which involves second neighbor hoppings, are the dimensionless extrinsic SO scattering strength $\lambda_{\rm SO}=\lambda \hbar/(4a^2)$, $\epsilon_{ijz}$ stands for the Levi-Civita totally antisymmetric tensor with $i,j$ denoting the in-plane coordinate axes, and $\nu,\gamma$ are the dummy indices taking values $\pm 1$. For 2DEG of Ref.~\cite{sih2005a} these SO coupling parameters take values $t_{\rm SO} \simeq 0.003t_{\rm o}$ and $\lambda_{\rm SO} \simeq 0.005$, assuming conduction bandwidth $\simeq 1$ eV and using the effective mass $m^*=0.074m_0$~\cite{sih2005a}.

{\it Mesoscopic fluctuations and scaling of the spin Hall conductance}.---Akin to the dichotomy in 
the description of the quantum Hall effect~\cite{ando2003}, the characterization of the spin 
Hall effect in terms of $\sigma_{sH}$ is (assuming ``reliable'' definition of spin current~\cite{niu2006a}) applicable at sufficiently high temperatures ensuring the classical regime where dephasing length $L_\phi$ is much smaller than the sample size and conductivity can be treated as a local quantity. On 
the other hand, in mesoscopic samples of size $L < L_\phi$ nonlocal quantum corrections to semiclassical background and presence of external measuring circuit have to be taken into account through $G_{sH}$. 

For example, quantum coherence and stochasticity introduced by the impurities will lead to sample-to-sample  fluctuations of $G_{sH}$, so that lack of self-averaging requires to study its full distribution 
function $P(G_{sH})$, as demonstrated by Fig.~\ref{fig:spin_ucf}. Nevertheless, the main part of the distribution is best characterized by the typical (i.e., geometric mean)~\cite{dobro2003a} conductance in Fig.~\ref{fig:spin_ucf}, rather than the usual arithmetic mean value affected by the long tails of $P(G_{sH})$. The inset of Fig.~\ref{fig:spin_ucf} also reveals sample-to-sample fluctuations of the 
voltages that have to be applied to transverse leads to ensure pure spin Hall current $I_2^{S_z}=-I_3^{S_z}$ (note that extrinsic SO scattering generates only the $z$-polarized spin Hall current studied here, while Rashba SO coupling is also responsible for non-zero spin currents with in-plane polarized spins~\cite{nikolic_mesoshe}) with no net charge current $I_2=I_3=0$ in the transverse direction. 
In perfectly clean and symmetric bars this condition is satisfied by applying   $V_2=V_3=(V_1-V_4)/2$~\cite{nikolic_mesoshe}.

\begin{figure}
\centerline{\psfig{file=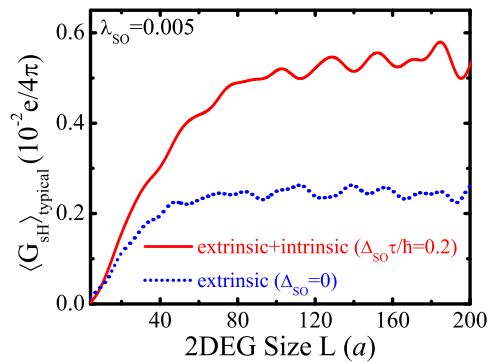,scale=0.6,angle=0}} 
\caption{(Color online) Scaling of the typical spin Hall 
conductance  with the size of the 2DEG characterized 
by the  extrinsic SO scattering strength $\lambda_{\rm SO}=0.005$ and the intrinsic 
Rashba SO coupling $t_{\rm SO}=0.003t_{\rm o}$ which, together with the disorder setting the mean 
free path $\ell =v_F \tau = 34a$, determine $\Delta_{\rm SO} \tau/\hbar=0.2$. For comparison, 
the ``extrinsic'' curve corresponds to $t_{\rm SO}=0=\Delta_{\rm SO}$.}\label{fig:scaling}
\end{figure}

Although the samples in Fig.~\ref{fig:scaling}, showing the scaling of $\langle G_{sH} \rangle (L)$,  are of the size $L \lesssim 1 \ \mu$m (compared to $100 \, \mu$m sizes of samples required for optical scanning of the spin Hall accumulation~\cite{kato2004a,sih2005a}), their properties can be used to account for spin Hall quantities of microstructures viewed  as a classical network of resistors, each of the size $L_\phi \times L_\phi$. The conductance $\langle G_{sH} \rangle (L=L_\phi)$ of each resistor can be read from Fig.~\ref{fig:scaling}, as computed in fully phase-coherent regime with localization length $\xi \gg L > L_\phi$. Even though inelastic scattering will destroy quantum-coherent nature  of transport between resistors, thereby pushing it into the semiclassical regime~\cite{sinova2004a,inoue2004a}, the spin 
Hall conductance of the whole network (i.e., of a macroscopic 2DEG bar at high temperatures) is 
still equal to $\langle G_{sH} \rangle (L=L_\phi)$ in two-dimensions.

{\it Conclusions}.---Our principal result in Fig.~\ref{fig:she_intro} demonstrates that the spin Hall effect 
in four-terminal disordered 2DEGs can be affected by both the SO scattering of impurities and the Rashba SO coupling inducing the energy spin-splitting $\Delta_{\rm SO}$. The former dominates $G_{sH} \simeq G_{sH}^{\rm extr}$ in the dirty regime $\Delta_{\rm SO} \tau/\hbar \lesssim 10^{-1}$, while the latter accounts for $G_{sH} \simeq G_{sH}^{\rm intr}$ in the clean regime $\Delta_{\rm SO}\tau/\hbar \gtrsim 1$ where it could be tuned via the gate electrode~\cite{nitta1997a} to two orders of magnitude larger value (for presently achievable Rashba coupling strengths $t_{\rm SO} \sim 0.01$~\cite{nitta1997a}) than the maximum value of $G_{sH}^{\rm extr}$ reached in the dirty limit. In the crossover regime $10^{-1} \lesssim \Delta_{\rm SO} \tau/\hbar \lesssim 1$, our nonperturbative quantum transport analysis unveils an interplay of the extrinsic and intrinsic mechanisms responsible for $G_{sH} \neq G_{sH}^{\rm extr} + G_{sH}^{\rm intr}$. These predictions offer a clear guidance toward nanofabrication of high-mobility devices in order to increase the magnitude of the spin Hall current, as well as to evade large sample-to-sample fluctuations of $G_{sH}$ and its sign change in Fig.~\ref{fig:spin_ucf} characterizing the low-temperature quantum spin Hall transport in  dirty phase-coherent bars.

\begin{acknowledgments}
We are grateful to  B.~I.~Halperin, N.~Nagaosa, and K.~Nomura for inspiring discussions. This research was supported in part by ACS grant No. PRF-41331-G10.
\end{acknowledgments}

\end{document}